\documentclass[12pt]{article}
\usepackage{lineno}
\usepackage{epsfig}
\usepackage{amsmath}
\usepackage{hhline}
\usepackage{amssymb}
\usepackage{times}
\usepackage{cite}
\usepackage{rotating}
\usepackage{url}

\newlength{\dinwidth}
\newlength{\dinmargin}
\def\EPJC{{Eur. Phys. J.} {\bf C}}

\def\be{\begin{equation}}
\def\ee{\end{equation}}
\def\bea{\begin{eqnarray}}
\def\eea{\end{eqnarray}}
\def\etal{{\it et~al.}}

\begin{document}  
\newcommand{\pom}{{I\!\!P}}
\newcommand{\slowpi}{\pi_{\mathit{slow}}}
\newcommand{\fiidiii}{F_2^{D(3)}}
\newcommand{\fiidiiiarg}{\fiidiii\,(\beta,\,Q^2,\,x)}
\newcommand{\n}{1.19\pm 0.06 (stat.) \pm0.07 (syst.)}
\newcommand{\nz}{1.30\pm 0.08 (stat.)^{+0.08}_{-0.14} (syst.)}
\newcommand{\fiidiiiful}{F_2^{D(4)}\,(\beta,\,Q^2,\,x,\,t)}
\newcommand{\fiipom}{\tilde F_2^D}
\newcommand{\ALPHA}{1.10\pm0.03 (stat.) \pm0.04 (syst.)}
\newcommand{\ALPHAZ}{1.15\pm0.04 (stat.)^{+0.04}_{-0.07} (syst.)}
\newcommand{\fiipomarg}{\fiipom\,(\beta,\,Q^2)}
\newcommand{\pomflux}{f_{\pom / p}}
\newcommand{\nxpom}{1.19\pm 0.06 (stat.) \pm0.07 (syst.)}
\newcommand {\gapprox}
   {\raisebox{-0.7ex}{$\stackrel {\textstyle>}{\sim}$}}
\newcommand {\lapprox}
   {\raisebox{-0.7ex}{$\stackrel {\textstyle<}{\sim}$}}
\def\gsim{\,\lower.25ex\hbox{$\scriptstyle\sim$}\kern-1.30ex%
\raise 0.55ex\hbox{$\scriptstyle >$}\,}
\def\lsim{\,\lower.25ex\hbox{$\scriptstyle\sim$}\kern-1.30ex%
\raise 0.55ex\hbox{$\scriptstyle <$}\,}
\newcommand{\pomfluxarg}{f_{\pom / p}\,(x_\pom)}
\newcommand{\dsf}{\mbox{$F_2^{D(3)}$}}
\newcommand{\dsfva}{\mbox{$F_2^{D(3)}(\beta,Q^2,x_{I\!\!P})$}}
\newcommand{\dsfvb}{\mbox{$F_2^{D(3)}(\beta,Q^2,x)$}}
\newcommand{\dsfpom}{$F_2^{I\!\!P}$}
\newcommand{\gap}{\stackrel{>}{\sim}}
\newcommand{\lap}{\stackrel{<}{\sim}}
\newcommand{\fem}{$F_2^{em}$}
\newcommand{\tsnmp}{$\tilde{\sigma}_{NC}(e^{\mp})$}
\newcommand{\tsnm}{$\tilde{\sigma}_{NC}(e^-)$}
\newcommand{\tsnp}{$\tilde{\sigma}_{NC}(e^+)$}
\newcommand{\st}{$\star$}
\newcommand{\sst}{$\star \star$}
\newcommand{\ssst}{$\star \star \star$}
\newcommand{\sssst}{$\star \star \star \star$}

\newcommand{\tw}{\theta_W}
\newcommand{\sw}{\sin{\theta_W}}
\newcommand{\cw}{\cos{\theta_W}}
\newcommand{\sww}{\sin^2{\theta_W}}
\newcommand{\cww}{\cos^2{\theta_W}}
\newcommand{\trm}{m_{\perp}}
\newcommand{\trp}{p_{\perp}}
\newcommand{\trmm}{m_{\perp}^2}
\newcommand{\trpp}{p_{\perp}^2}
\newcommand{\alp}{\alpha_s}

\newcommand{\alps}{\alpha_s}
\newcommand{\sqrts}{$\sqrt{s}$}
\newcommand{\LO}{$O(\alpha_s^0)$}
\newcommand{\Oa}{$O(\alpha_s)$}
\newcommand{\Oaa}{$O(\alpha_s^2)$}
\newcommand{\PT}{p_{\perp}}
\newcommand{\JPSI}{J/\psi}
\newcommand{\sh}{\hat{s}}
\newcommand{\uh}{\hat{u}}
\newcommand{\MP}{m_{J/\psi}}
\newcommand{\PO}{I\!\!P}
\newcommand{\xbj}{x}
\newcommand{\xpom}{x_{\PO}}
\newcommand{\ttbs}{\char'134}
\newcommand{\xpomlo}{3\times10^{-4}}  
\newcommand{\xpomup}{0.05}  
\newcommand{\dgr}{^\circ}
\newcommand{\pbarnt}{\,\mbox{{\rm pb$^{-1}$}}}
\newcommand{\gev}{\,\mbox{GeV}}
\newcommand{\WBoson}{\mbox{$W$}}
\newcommand{\fbarn}{\,\mbox{{\rm fb}}}
\newcommand{\fbarnt}{\,\mbox{{\rm fb$^{-1}$}}}
%
%
\newcommand{\qsq}{\ensuremath{Q^2} }
\newcommand{\gevsq}{\ensuremath{\mathrm{GeV}^2} }
\newcommand{\et}{\ensuremath{E_t^*} }
\newcommand{\rap}{\ensuremath{\eta^*} }
\newcommand{\gp}{\ensuremath{\gamma^*}p }
\newcommand{\dsiget}{\ensuremath{{\rm d}\sigma_{ep}/{\rm d}E_t^*} }
\newcommand{\dsigrap}{\ensuremath{{\rm d}\sigma_{ep}/{\rm d}\eta^*} }
\newcommand{\MSbar}{\ensuremath{\overline{\text{MS}}\ }}

\begin{titlepage}

\noindent
\begin{flushleft}
{\tt DESY-17-048} \\
{\tt 28 April 2017} \\

\end{flushleft}


\vspace{2.5cm}
\begin{center}
\begin{Large}
\boldmath
{\bf Running of the Charm-Quark Mass from HERA \\
Deep-Inelastic Scattering Data}
\unboldmath

\end{Large}
\end{center}

\vspace{1cm}
\def\gev{\rm GeV}
\def\ie{\it i.e.}
\def\etal{\hbox{$\it et~al.$}}
\def\clb#1 {(#1 Coll.),}
\hyphenation{do-mi-nant
}

\begin{abstract}
\noindent

Combined HERA data on charm production in deep-inelastic scattering
have previously been used to determine the charm-quark running mass 
$m_c(m_c)$
in the \MSbar re\-nor\-ma\-li\-sa\-tion scheme. 
Here, the same data are used
as a function of the photon virtuality $Q^2$ to evaluate the charm-quark
running mass at different scales to one-loop order, in the context of 
a next-to-leading order QCD analysis.
The scale dependence of the mass is found to be consistent with QCD
expectations.

\end{abstract}
\vspace{1cm}

{\small\raggedright
A.~Gizhko$^{\mathrm{9}}$,
A.~Geiser$^{\mathrm{9}}$,
S.~Moch$^{\mathrm{7}}$,
I.~Abt$^{\mathrm{16}}$,
O.~Behnke$^{\mathrm{9}}$,
A.~Bertolin$^{\mathrm{19}}$,
J. Bl\"umlein$^{\mathrm{26}}$,
D.~Britzger$^{\mathrm{9}}$,
R.~Brugnera$^{\mathrm{20}}$,
A.~Buniatyan$^{\mathrm{2}}$,
P.J.~Bussey$^{\mathrm{6}}$,
R.~Carlin$^{\mathrm{20}}$,
A.M.~Cooper-Sarkar$^{\mathrm{18}}$,
K.~Daum$^{\mathrm{25}}$,
S.~Dusini$^{\mathrm{19}}$,
E.~Elsen$^{\mathrm{9},\mathrm{a3}}$,
L.~Favart$^{\mathrm{1}}$,
J.~Feltesse$^{\mathrm{5}}$,
B.~Foster$^{\mathrm{8},\mathrm{a1}}$,
A.~Garfagnini$^{\mathrm{20}}$,
M.~Garzelli$^{\mathrm{7}}$,
J.~Gayler$^{\mathrm{9}}$,
D.~Haidt$^{\mathrm{9}}$,
J.~Hladk\`y$^{\mathrm{21}}$,
A.W.~Jung$^{\mathrm{10},\mathrm{a2}}$,
M.~Kapichine$^{\mathrm{4}}$,
I.A.~Korzhavina$^{\mathrm{15}}$,
B.B.~Levchenko$^{\mathrm{15}}$,
K.~Lipka$^{\mathrm{9}}$,
M.~Lisovyi$^{\mathrm{9},\mathrm{a5}}$,
A.~Longhin$^{\mathrm{19}}$,
S.~Mikocki$^{\mathrm{12}}$,
Th.~Naumann$^{\mathrm{26}}$,
G.~Nowak$^{\mathrm{12}}$,
E.~Paul$^{\mathrm{3}}$,
R.~Pla\v{c}akyt\.{e}$^{\mathrm{7}}$,
K.~Rabbertz$^{\mathrm{11}}$,
S.~Schmitt$^{\mathrm{9}}$,
L.M.~Shcheglova$^{\mathrm{15},\mathrm{a6}}$,
Z.~Si$^{\mathrm{22}}$,
H.~Spiesberger$^{\mathrm{14}}$,
L.~Stanco$^{\mathrm{19}}$,
P.~Tru\"ol$^{\mathrm{27}}$,
T.~Tymieniecka$^{\mathrm{24}}$,
A.~Verbytskyi$^{\mathrm{16}}$,
K.~Wichmann$^{\mathrm{9},\mathrm{b5}}$,
M.~Wing$^{\mathrm{13},\mathrm{a4}}$,
A.F.~\.Zarnecki$^{\mathrm{23}}$,
O.~Zenaiev$^{\mathrm{9}}$,
Z.~Zhang$^{\mathrm{17}}$

\bigskip
\footnotesize\begin{description}\setlength{\parsep}{0em}\setlength{\itemsep}{0em}
\item[$^{1}$] 
 Inter-University Institute for High Energies ULB-VUB, Brussels and Universiteit Antwerpen, Antwerpen, Belgium$^{\mathrm{b1}}$
\item[$^{2}$] 
 School of Physics and Astronomy, University of Birmingham, Birmingham, UK$^{\mathrm{b2}}$
\item[$^{3}$] 
 {Physikalisches Institut der Universit\"at Bonn, Bonn, Germany{}}$^{\mathrm{b4}}$
\item[$^{4}$] 
 Joint Institute for Nuclear Research, Dubna, Russia
\item[$^{5}$] 
 Irfu/SPP, CE-Saclay, Gif-sur-Yvette, France
\item[$^{6}$] 
 {School of Physics and Astronomy, University of Glasgow, Glasgow, United Kingdom{}}$^{\mathrm{b2}}$
\item[$^{7}$] 
 II. Institute for Theoretical Physics, Hamburg University, Hamburg, Germany
\item[$^{8}$] 
 Institut f\"ur Experimentalphysik, Universit\"at Hamburg, Hamburg, Germany$^{\mathrm{b6},\mathrm{b7}}$
\item[$^{9}$] 
 {Deutsches Elektronen-Synchrotron DESY, Hamburg, Germany}
\item[$^{10}$] 
 Kirchhoff-Institut f\"ur Physik, Universit\"at Heidelberg, Heidelberg, Germany$^{\mathrm{b6}}$
\item[$^{11}$] 
 Karlsruher Institut f\"ur Technologie, Karlsruhe, Germany
\item[$^{12}$] 
 {The Henryk Niewodniczanski Institute of Nuclear Physics, Polish Academy of \ Sciences, Krakow, Poland{}}$^{\mathrm{b9}}$
\item[$^{13}$] 
 {Physics and Astronomy Department, University College London, London, United Kingdom{}}$^{\mathrm{b2}}$
\item[$^{14}$] 
 PRISMA Cluster of Excellence, Institut fur Physik,
Johannes Gutenberg-Universit\"at, Mainz, Germany,
and Centre for Theoretical and Mathematical Physics and Department of Physics,
University of Cape Town, Rondebosch 7700, South Africa
\item[$^{15}$] 
 {Lomonosov Moscow State University, Skobeltsyn Institute of Nuclear Physics, Moscow, Russia{}}$^{\mathrm{b8}}$
\item[$^{16}$] 
 {Max-Planck-Institut f\"ur Physik, M\"unchen, Germany}
\item[$^{17}$] 
 LAL, Universit\'e Paris-Sud, CNRS/IN2P3, Orsay, France
\item[$^{18}$] 
 {Department of Physics, University of Oxford, Oxford, United Kingdom{}}$^{\mathrm{b2}}$
\item[$^{19}$] 
 {INFN Padova, Padova, Italy{}}$^{\mathrm{b3}}$
\item[$^{20}$] 
 {Dipartimento di Fisica e Astronomia dell' Universit\`a and INFN, Padova, Italy{}}$^{\mathrm{b3}}$
\item[$^{21}$] 
 Institute of Physics, Academy of Sciences of the Czech Republic, Praha, Czech Republic$^{\mathrm{b10}}$
\item[$^{22}$] 
Shandong University, Jinan, Shandong Province, P.R. China
\item[$^{23}$] 
 {Faculty of Physics, University of Warsaw, Warsaw, Poland}
\item[$^{24}$] 
 {National Centre for Nuclear Research, Warsaw, Poland}
\item[$^{25}$] 
 Fachbereich C, Universit\"at Wuppertal, Wuppertal, Germany
\item[$^{26}$] 
 {Deutsches Elektronen-Synchrotron DESY, Zeuthen, Germany}
\item[$^{27}$] 
 Physik-Institut der Universit\"at Z\"urich, Z\"urich, Switzerland$^{\mathrm{b11}}$
\end{description}
\medskip\goodbreak
\begin{description}\setlength{\parsep}{0em}\setlength{\itemsep}{0em}
\item[$^{\mathrm{a1}}$] %
 Alexander von Humboldt Professor; also at DESY and University of Oxford
\item[$^{\mathrm{a2}}$] %
 Now at Fermilab, Chicago, USA, and
 Purdue University, West Lafayette, USA
\item[$^{\mathrm{a3}}$] %
 Now at CERN, Geneva, Switzerland
\item[$^{\mathrm{a4}}$] %
 Also supported by DESY and the Alexander von Humboldt Foundation
\item[$^{\mathrm{a5}}$] %
Now at Physikalisches Institut, Universit\"{a}t Heidelberg, Germany
\item[$^{\mathrm{a6}}$] %
Also at University of Bristol, UK.
\end{description}
\medskip\goodbreak
\begin{description}\setlength{\parsep}{0em}\setlength{\itemsep}{0em}
\item[$^{\mathrm{b1}}$] %
 Supported by FNRS-FWO-Vlaanderen, IISN-IIKW and IWT and by Interuniversity Attraction Poles Programme, Belgian Science Policy
\item[$^{\mathrm{b2}}$] %
 Supported by the UK Science and Technology Facilities Council
\item[$^{\mathrm{b3}}$] %
 Supported by the Italian National Institute for Nuclear Physics (INFN)
\item[$^{\mathrm{b4}}$] %
 Supported by the German Federal Ministry for Education and Research (BMBF), under contract No. 05 H09PDF
\item[$^{\mathrm{b5}}$] %
 Supported by the Alexander von Humboldt Foundation
\item[$^{\mathrm{b6}}$] %
 Supported by the Bundesministerium f\"ur Bildung und Forschung, FRG, under contract number 05H09GUF
\item[$^{\mathrm{b7}}$] %
 Supported by the SFB 676 of the Deutsche Forschungsgemeinschaft (DFG)
\item[$^{\mathrm{b8}}$] %
 Partially Supported by RF Presidential grant NSh-7989.2016.2
\item[$^{\mathrm{b9}}$] %
 Partially Supported by Polish Ministry of Science and Higher Education, grant DPN/N168/DESY/2009
\item[$^{\mathrm{b10}}$] %
 Supported by the Ministry of Education of the Czech Republic under the project INGO-LG14033
\item[$^{\mathrm{b11}}$] %
 Supported by the Swiss National Science Foundation
\end{description}}

\vspace{2cm}

\end{titlepage}

\newpage
\section{Introduction}

The Standard Model of particle physics is based on Quantum Field Theory,
which can provide predictions that rely on a perturbative approach.
In the \MSbar re\-nor\-ma\-li\-sa\-tion 
scheme of perturbative quantum chromodynamics 
(pQCD), the values of all basic QCD parameters depend on the scale $\mu$ 
at which they are evaluated.
The most prominent example is the scale dependence, i.e. running, of the 
strong 
coupling constant $\alpha_s$, a by now well established property of pQCD. 
It has, for example, been determined from measurements of hadronic event shapes
or jet production at $e^+e^-$ colliders \cite{PETRA, LEPalpharun}, and from 
measurements of jet production at HERA \cite{HERAalpharun}, 
Tevatron \cite{Tevaalpharun} and LHC \cite{LHCalpharun}. 

The scale dependence of the mass $m_Q$ of a heavy quark in the \MSbar scheme can likewise be 
evaluated perturbatively, using the renormalisation group equation 
\begin{eqnarray}
  \label{eq:rundec}
  \mu^2\,\frac{d}{d\mu^2}m_Q(\mu) &=& m_Q(\mu)\,\gamma_{m_Q}\left(\alpha_s\right) 
  \, ,
\end{eqnarray}
which is governed by the mass anomalous dimension $\gamma_{m_Q}(\alpha_s)$ 
known up to five-loop order~\cite{Baikov:2014qja} in perturbation theory.
The running of the \MSbar beauty-quark mass has already been successfully 
investigated from measurements at the LEP $e^+ e^-$ collider \cite{beautyrun}.
Heavy-flavour production in deep-inelastic scattering (DIS) at HERA is 
particularly sensitive to heavy-quark pair production at the kinematic 
threshold. 
A recent determination of the beauty-quark mass $m_b(m_b)$ \cite{ZEUSbeauty} 
by the ZEUS experiment at HERA was reinterpreted as a measurement of 
$m_b(\mu=2m_b)$ using the solution of Eq.~(\ref{eq:rundec}) at one loop. 
The comparison \cite{Addendum,Andriithesis,HQreview} of this result with the 
measurements from LEP and the PDG world average \cite{PDG12,PDG16} shows 
consistency with the expected running of the beauty-quark mass. 
  
\begin{figure}[ht]
\center
\epsfig{file=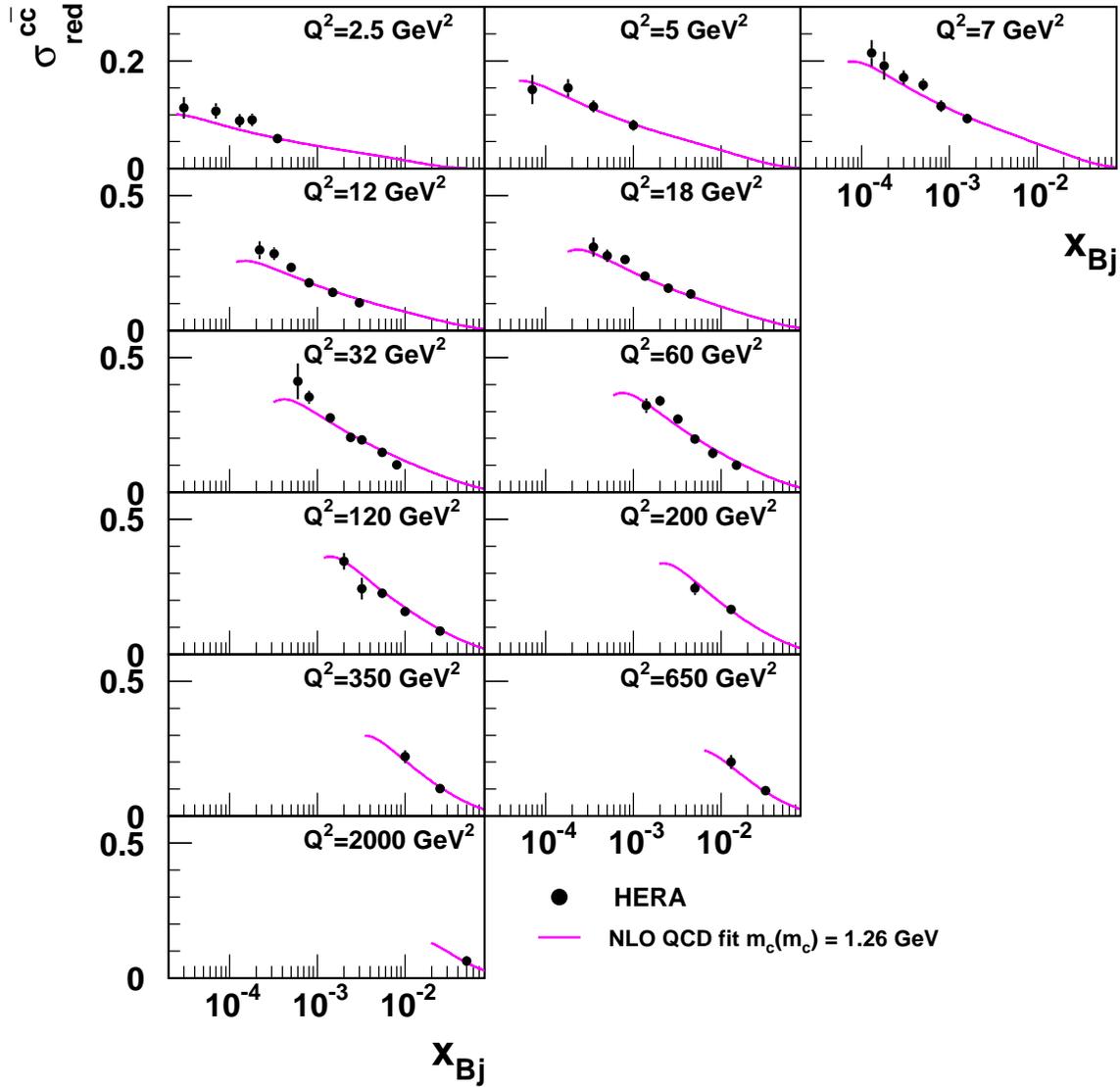,width=1.0\textwidth}
\caption{
Reduced cross section for charm production in deep-inelastic 
scattering \cite{HERAcharmcomb} as a function of the Bjorken scaling variable 
$x_{\rm Bj}$ for different values of photon virtuality $Q^2$ (points). 
The measurements are grouped into six subsets in $Q^2$, as indicated by the
six rows, and detailed in Table \ref{tab:runval}. The curve shows the 
global NLO QCD fit for $m_c(m_c)=1.26$ GeV described in the text. 
} 
\label{fig:sigred} 
\end{figure}

An explicit investigation of the running of the charm-quark mass has not 
been performed yet.
Combined HERA measurements \cite{HERAcharmcomb} on charm 
production in deep-inelastic scattering
have already been used for several determinations of the 
charm-quark mass $m_c(\mu=m_c)$ in the \MSbar renormalisation scheme 
\cite{HERAcharmcomb,adlm2,CTEQmc,FONLLmc,ABMP16}.
Figure \ref{fig:sigred} shows the measured reduced cross section for charm 
production \cite{HERAcharmcomb} as a 
function of the Bjorken variable $x_{\rm Bj}$ in 12 bins of photon virtuality 
$Q^2$ in the range $2.5$ GeV$^2<Q^2<2000$ GeV$^2$. 
In this paper, these data 
are used to investigate the running of the charm-quark mass 
with the same treatment 
of the uncertainties of the combination as in Ref. \cite{HERAcharmcomb}.
The fixed flavour number scheme (FFNS) is used at 
next-to-leading order (NLO) with $n_f=3$ active flavours. This scheme 
gives a very good description of the charm data 
\cite{HERAcharmcomb,Alekhin:2013nda}, as shown in Fig. \ref{fig:sigred}. 
Calculations of next-to-next-to-leading order corrections with massive 
coefficient functions \cite{ABMP16,Alekhin:2013nda} have not yet been 
completed, and are therefore not used in this paper.

\section{Principle of the $m_c(\mu)$ determination}
\label{sect:principle}

The theoretical reduced cross section for charm production is obtained from a 
convolution 
of charm-production matrix elements with appropriate parton density functions 
(PDFs). The latter are obtained from inclusive DIS cross sections, which 
include a charm contribution. Thus both, matrix elements and PDFs, depend on 
the value of the charm-quark mass. 
The scale dependence of the charm-quark mass is evaluated 
by subdividing the charm cross-section data \cite{HERAcharmcomb} into several 
subsets corresponding to different individual scales, as indicated by 
different rows in Fig.~\ref{fig:sigred}. 
In contrast, in the evaluation of 
the PDFs, data spanning a large scale range such as the inclusive HERA 
DIS data \cite{DIScomb,DIScombII} must be used in order to get significant 
PDF constraints. A subdivision into individual scale ranges is thus not 
possible for the PDF determination. On the other hand, 
it has been established that, apart from the strong constraint which the
charm measurements impose on the charm-quark mass \cite{HERAcharmcomb}, 
their influence on a combined PDF fit of both inclusive and 
charm data is small \cite{DIScombII}. Therefore, the PDFs extracted from 
inclusive DIS can be used for investigations of charm-quark properties,
provided that the same charm-quark mass is used throughout, recognising
that thereby some correlation between the mass and PDF extractions is 
induced. 
The influence of this correlation on the determination
of the charm-quark mass running is minimised as described in 
section \ref{sect:Results}.
  
To obtain the charm-quark mass at different scales,
the charm data are 
subdivided into six kinematic intervals according to the virtuality of 
the exchanged photon.
Each measurement in a given range in $Q^2$, as listed in Table~\ref{tab:runval} 
and shown in Fig.~\ref{fig:sigred}, 
is performed with charm data originating from collisions at a typical 
scale of $\mu = \sqrt{Q^2 + 4m_c^2}$. The actual scale used for each interval
is defined according to 
\begin{equation} \label{eq:scale}
\log \mu = \left< \log\left(\sqrt{Q^2 + 4m_c^2}\right)\right>,
\end{equation}
where the brackets indicate the logarithmic average of the considered range. 
The resulting value for each $Q^2$ range is also 
listed in Table~\ref{tab:runval}.

Technically, a value of $m_c(m_c)$ is extracted separately from a fit to 
each interval. 
The value of $m_c(m_c)$ is obtained assuming the running of both $\alpha_s$ and 
$m_c$ as predicted by QCD. 
%
%
To that end,  Eq.~(\ref{eq:rundec}) is solved using the one-loop dependence on the scale $\mu$, 
as relevant in a NLO calculation, as
\begin{equation} 
  \label{eq:running}
  m_{Q}(\mu)\,=\,m_{Q}(m_{Q})\times\left(\frac{\alpha_{s}(\mu)}{\alpha_{s}(m_{Q})}\right)^{c_0}, 
\end{equation}
where $c_0= 4/(11-2n_f/3) = 4/9$ as appropriate for QCD with $n_f=3$ for the 
number of light quark flavours. 
Equation (\ref{eq:running}) is used to evaluate the mass running in
all results of this work. 

Expanded and truncated to leading order in powers of $\alpha_s$, 
this can also be expressed in the form (not used here)
\begin{equation} 
  m_{Q}(\mu) = m_{Q}(m_Q)\left(1 + \frac{\alpha_s(\mu)}{\pi} \log\left(\frac{\mu^2}{m_Q^2}\right) + O(\alpha_s^2) \right)
  \, . 
\end{equation}
This illustrates that the scale dependence is logarithmic and justifies the 
logarithmic average in Eq.~(\ref{eq:scale}).

According to Eq.~(\ref{eq:running}) the mass has actually been determined at 
the scale $\mu$, and was extrapolated to the scale $m_c$ when expressed 
as $m_c(m_c)$.
If each determination of $m_c(m_c)$ is reinterpreted in terms of a value of 
$m_c(\mu)$ using Eq.~(\ref{eq:running}), 
the mass determinations are reverted to their unextrapolated value, 
and the effect of the initial 
assumption of QCD running on the interpretation of their value is 
minimised for the final result.

\section{QCD predictions and systematic uncertainties}
\label{sect:theory}

QCD predictions for the reduced charm cross sections are obtained at 
NLO in pQCD ($O(\alpha_s^2)$) using the OPENQCDRAD 
package \cite{openqcdrad} as available in HERAFitter\footnote{Recently renamed
xFitter.} 
\cite{herafitter,DIScomb}.
These predictions are based on the ABM implementation \cite{abkm09msbar} 
of charm cross-section calculations
in the 3-flavour FFNS.
The renormalisation and factorisation scales are always taken to be identical.
In the calculations, the same settings and parametrisations are chosen as 
those used for the earlier measurement of $m_c(m_c)$ \cite{HERAcharmcomb}. 
In addition, scale variations were applied as in Ref. \cite{adlm2}. 
For all explicit calculations of charm-quark mass running, an implementation 
of the one-loop formula~\cite{RUNdec}, Eq.~(\ref{eq:running}), is used, 
which is consistent with that used implicitly in OPENQCDRAD.

These predictions are fitted to the data. 
The fit uncertainty $\delta^{\rm exp}_{\rm fit}$ is determined by applying the 
criterion $\triangle \chi^{2} = 1$ with the same formalism as in 
Ref. \cite{HERAcharmcomb}. It contains 
the experimental uncertainties, the extrapolation uncertainties and
the uncertainties of the default PDF parametrisation. In addition, 
the result has uncertainties attributed to the choices of extra model 
parameters, additional variations of the PDF parametrisation and 
uncertainties on the perturbative QCD parameters as listed in terms of 
$\delta_1$ to $\delta_7$ below.

The following additional parameters are used in the calculations, presented 
with the
variations performed to estimate their systematic uncertainties 
\begin{itemize}
\item {$\delta_{1}$: \bf \MSbar running mass of the beauty quark}, 
$m_b(m_b)=4.75$~GeV, varied within the range  
$m_b(m_b)=4.3$ GeV to $m_b(m_b)=5.0$ GeV, 
to be consistent with  \cite{HERAcharmcomb};
\item {$\delta_{2}$: \bf strong coupling constant} $\alpha_s^{n_f=3,{\rm NLO}}(M_Z) = 0.105 \pm 0.002$,
 corresponding to \\
 $\alpha_s^{n_f=5, {\rm NLO}}(M_Z) = 0.116 \pm 0.002$, as in \cite{HERAcharmcomb};
\item {$\delta_{3}$: \bf strangeness suppression factor} $f_{s}=0.31$, 
varied within the range $f_{s}=0.23$ to $f_{s}=0.38$, 
as in \cite{HERAcharmcomb};
\item {$\delta_{4}$: \bf evolution starting scale} 
$Q^{2}_{0}=1.4$~GeV$^{2}$, varied to $Q^{2}_{0}=1.9$~GeV$^{2}$, 
as in \cite{HERAcharmcomb};
\item {$\delta_{5}$: \bf minimum $Q^{2}$ of inclusive data in the fit $Q^{2}_{\rm min}$}.
 For the PDF extraction, the minimum $Q^{2}$ of the inclusive data was set to 
$Q^{2}_{\rm min}=3.5$~GeV$^{2}$ and varied to $Q^{2}_{\rm min}=5$~GeV$^{2}$, 
as in \cite{HERAcharmcomb};
\item {$\delta_{6}$: \bf the parametrisation of the proton structure} is 
described by a series of {FFNS} variants 
of the HERAPDF1.0 PDF set~\cite{DIScomb} at NLO, evaluated for the respective 
charm-quark mass, for $\alpha_s^{n_f=3, {\rm NLO}}(M_Z) = 0.105 \pm 0.002$, 
consistent with $\delta_2$. 

The additional PDF parametrisation uncertainties are calculated according to 
the HERAPDF1.0  prescription~\cite{DIScomb}, by freeing three extra PDF 
parameters $D_{u_{v}},D_{\bar{D}}$ and $D_{\bar{U}}$ in the fit;
 \item {$\delta_{7}$: \bf renormalisation and factorisation scales}
  $\mu_f=\mu_r=\sqrt{Q^2+4m_Q^2(m_Q)}=\mu$, varied simultaneously up 
  (upper value) or down (lower value) by
  a factor of two for the massive quark (charm and beauty) parts of the 
calculation, as in \cite{adlm2}.
\end{itemize}

The numerical values for each bin are shown in Table \ref{tab:runsyst}. 
The dominant uncertainties are those 
arising from $\delta^{\rm exp}_{\rm fit}$, followed by those from the scale 
variations $\delta_7$.

\section{Results}
\label{sect:Results}
  
In order to minimise the correlated contribution from inclusive data 
to the charm-mass determinations, and in particular from the implicit 
charm-mass scale dependence therein, a set of PDFs in the 3-flavour FFNS 
is extracted from a QCD fit to inclusive DIS HERA data
\cite{DIScomb}. This extraction uses 
exactly the same setup as that used in a previous 
publication \cite{HERAcharmcomb}, but allows for different charm-quark masses. 
The charm-quark mass as a function of scale is then extracted from a fit 
to the charm data only. 
When this 
analysis was originally performed \cite{Andriithesis,HQreview}, the 
inclusive HERA II DIS 
data \cite{DIScombII} were not yet available. 
The use of the earlier inclusive data \cite{DIScomb} has been retained
for several reasons. Firstly, all systematic 
uncertainties can be treated exactly as in the corresponding previous 
global $m_c(m_c)$ determination \cite{HERAcharmcomb}. Secondly, the newer and 
more precise inclusive data are more strongly sensitive \cite{FONLLmc} 
to the assumed charm-quark mass and its running than the earlier 
inclusive data. 
This is actually counterproductive for the purpose of this paper in which the 
cross-correlations to the inclusive data, which cannot be subdivided into 
scale intervals, need to be minimised. Thirdly, 
the uncertainties on the determination of charm-quark mass running arising 
from the PDF uncertainties are already small (Table~\ref{tab:runsyst}) 
compared to other uncertainties. 
For the purpose of this paper, the conceptual advantage of minimising the 
mass-related correlations between the charm
and inclusive data sets therefore outweighs the potential gain from 
a higher PDF precision.

\begin{figure}[ht]
\center
\epsfig{file=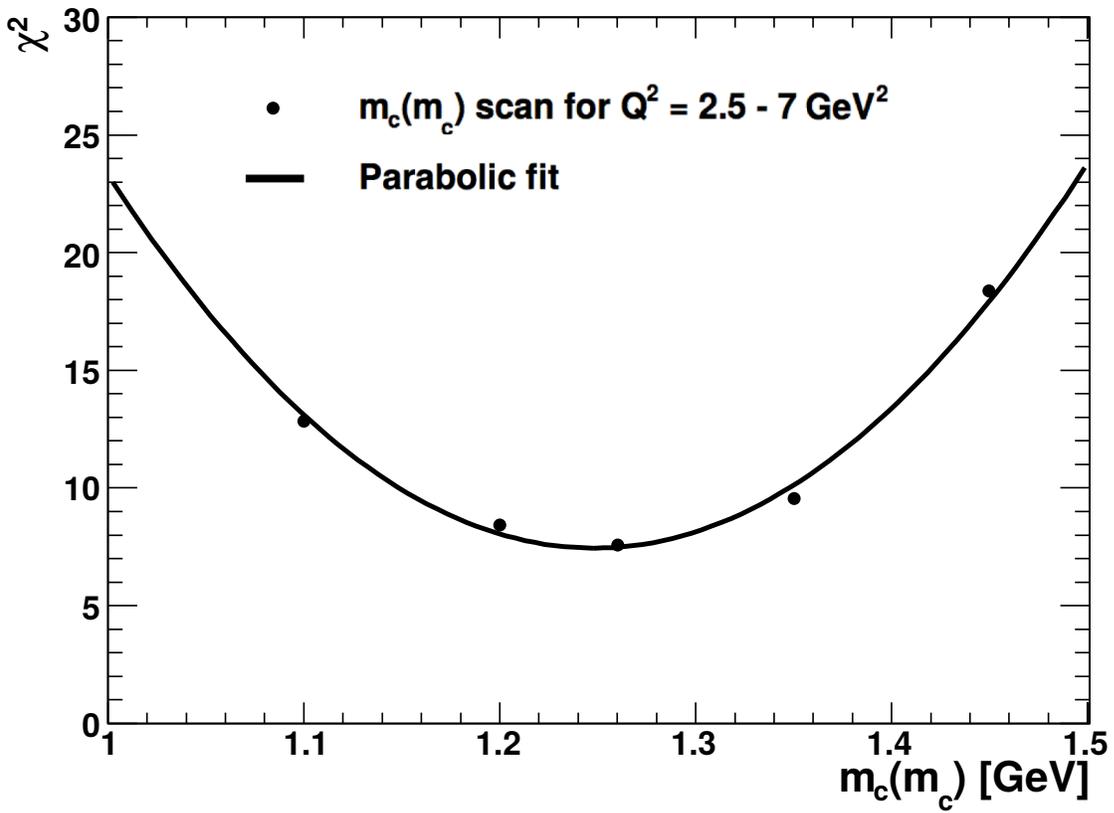,width=1.0\textwidth}
\caption{
$\chi^2$ of the comparison of the FFNS NLO QCD prediction to the 
charm reduced cross sections in the first $Q^2$ interval, $2.5-7$~GeV$^2$, for 
different values of the charm-quark mass $m_c(m_c)$ in the \MSbar running mass
scheme (points). The line shows a parabolic fit.
}
\label{fig:chi2scan} 
\end{figure}

For each charm-quark mass hypothesis, predictions for the reduced charm
cross sections are obtained using the corresponding PDF and are compared 
to one of the six subsets of the charm data listed
in Table \ref{tab:runval} and shown in Fig. \ref{fig:sigred}.
The \MSbar running mass of the charm quark is varied within the range
$m_c(m_c)=1$ GeV to $m_c(m_c)=1.5$ GeV in several steps. 
The $\chi^2$ distribution of this comparison 
is used to extract the value of the charm-quark mass $m_c(m_c)$. 
An example of such a distribution for the first $Q^2$ interval is shown in 
Fig.~\ref{fig:chi2scan}, together with a parabolic fit. 
The minimum yields the measured charm-quark mass, while the fit uncertainty is 
obtained from $\Delta\chi^2=1$. 
The corresponding distributions for the other 
intervals can be found in Ref. \cite{Andriithesis}. 
A global fit to the complete 
charm data set for $m_c(m_c) = 1.26$ GeV, the central value obtained from the 
earlier global $m_c(m_c)$ analysis \cite{HERAcharmcomb}, is  
shown as a curve in Fig. \ref{fig:sigred} for comparison. The data are well 
described.

\begin{figure}[ht]
\center
\epsfig{file=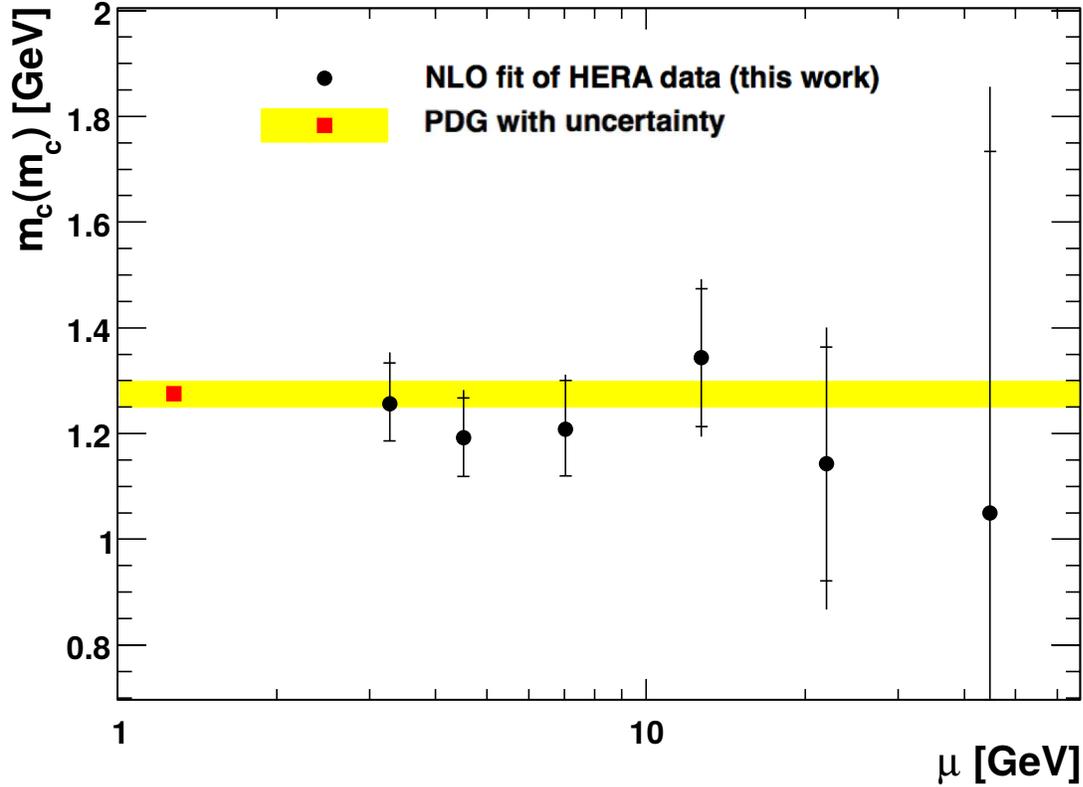,width=1.0\textwidth}
\caption{
Charm-quark mass $m_c(m_c)$ in the \MSbar running mass scheme 
determined from the charm data independently 
at six different scales $\mu$. 
The outer error bars show the fit uncertainty combined with all
model, parametrisation and theoretical systematic uncertainties added 
in quadrature. The inner error bars show the same uncertainties excluding 
the uncertainties arising from the variation of the QCD scales. 
The filled square at scale $m_c$ is the PDG world 
average \cite{PDG12} and the associated band shows its uncertainty.
}
\label{fig:mcmc} 
\end{figure}

The values of $m_c(m_c)$ extracted for each of the subsets of charm data 
are shown in Fig. \ref{fig:mcmc} and listed in 
Table \ref{tab:runval} as a function of the corresponding scale $\mu$, 
together with their uncertainties. 
The breakdown of the uncertainties into individual sources 
is summarised in Table \ref{tab:runsyst}. 
The values of $m_c(m_c)$ determined in the different subsets agree well within 
uncertainties with each other, with the value from the global analysis 
quoted above, and with the independent PDG world average\footnote{The 
PDG2012 \cite{PDG12} value is used 
since it does not yet contain the result from \cite{HERAcharmcomb} in the 
average, and is thus an independent value. The latest PDG2016 \cite{PDG16} 
value only differs very slightly from it.} of $1.275 \pm 0.025$ GeV
\cite{PDG12}.

In order to test the stability of the $m_c(m_c)$ determination, the default 
analysis procedure is cross-checked with an alternative method. 
For each $Q^2$ interval a simultaneous PDF fit of the charm data from this 
interval and the full inclusive DIS data is performed.
From the total $\chi^2$ obtained by these fits, the $\chi^2$ of the 
corresponding fits to the inclusive data only, is subtracted. 
These differences are then used  for the determination of $m_c(m_c)$ by a 
$\chi^2$ scan in the same way as for the standard procedure.  
Despite the more direct cross-correlation of the $\chi^2$ from the charm sample 
with that from the inclusive sample 
in this method, the difference between the results of both methods is found 
to be negligible \cite{Andriithesis}, i.e. smaller than the width of the line 
in Fig. \ref{fig:chi2scan}. This indicates that the 
residual effect of the cross-correlation is small.

In the final step, the values of $m_c(m_c)$ are consistently translated 
back to $m_c(\mu)$ assuming the running of $\alpha_s$ and $m_c$ as predicted 
by the QCD framework (Eq.~\ref{eq:running}).
The resulting values of $m_c(\mu)$ are included in Table \ref{tab:runval}. 
The fractional contributions of the uncertainties for the 
individual sources (before the translation) are the same as those 
for $m_c(m_c)$ as listed in Table 2, with the exception of the scale-variation
uncertainties $\delta_7$ as discussed below.

\begin{figure}[ht]
\center
\epsfig{file=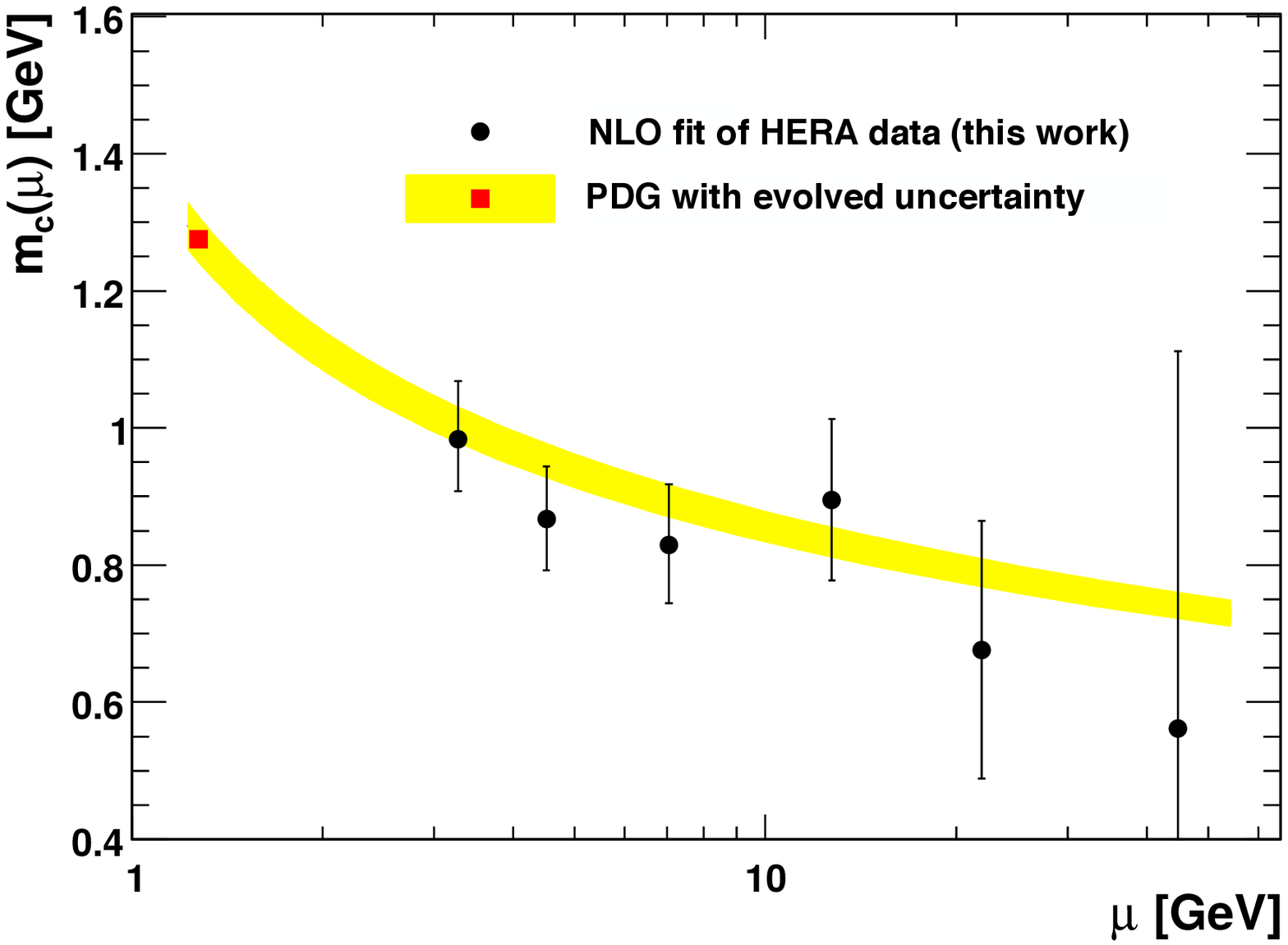,width=1.0\textwidth}
\caption{
Charm-quark mass $m_c(\mu)$ determined in the \MSbar running mass scheme
as a function of the scale $\mu$ 
(black points).
The error bars correspond to the inner error bars shown in Fig.~\ref{fig:mcmc}.
The red point at scale $m_c$ is the PDG world average \cite{PDG12} and the band 
shows the uncertainty and its expected running according to 
Eq.~(\ref{eq:running}).
}
\label{fig:mcrun} 
\end{figure}

In Fig. \ref{fig:mcrun}, the resulting scale dependence of $m_c(\mu)$ 
is shown together with the world average of $m_c(m_c)$ and the expectation for 
the evolution of $m_c(\mu)$ within the NLO QCD framework. 
The data are well described by the theoretical expectations. The running
of the charm-quark mass as a function of the scale $\mu$ is clearly visible, 
if the independent PDG point obtained mainly from low scale QCD lattice 
calculations is included.
 
No scale variations are shown in Fig. \ref{fig:mcrun}.
In addition to the scale uncertainties, $\delta_7$, extracted from the fit, 
these would correspond to a variation in scale of the horizontal axis of the 
figure and/or a shift of the points along the expected scale-dependence 
curve, which are difficult to represent graphically.
Furthermore, they are strongly correlated point by point, such that the shape 
of the distribution will stay essentially unchanged. 
In any case it is clear from 
Fig. \ref{fig:mcmc} that their effect is not dominant. 

Overall, this result is a nontrivial consistency 
check of the charm-quark mass running. It is conceptually similar to the 
procedure of extracting 
the running of $\alpha_s(\mu)$ from jet production at different 
transverse energy scales \cite{HERAalpharun, Tevaalpharun, LHCalpharun} 
or at different $e^+e^-$ centre-of-mass energies \cite{PETRA, LEPalpharun}.

\section{Conclusions}
\label{sect:conclude}

The running of the charm-quark mass $m_c(\mu)$ in the \MSbar scheme is 
evaluated for the first time,
using the combined reduced-cross-section charm data from HERA. 
It is found to be consistent with the expectation from QCD.
Within the limited scale range of each subset of the charm data used for the 
determination of $m_c(\mu)$, the running of the charm-quark mass is implicitly 
assumed as part of the QCD theory input. Therefore this determination is not 
fully unbiased. However, the implicit bias of each individual $m_c(\mu)$ value 
is much smaller than the bias of the earlier extractions of a single $m_c(m_c)$ 
value 
from the complete data set. Furthermore, the PDG 
value of $m_c(m_c)=1.275 \pm 0.025$ GeV indicated in Figs.
\ref{fig:mcmc} and \ref{fig:mcrun} is mainly 
obtained from lattice gauge theory and time-like processes at scales 
in the vicinity of the charm-quark mass, at which its value is displayed. 
Therefore the comparison to this independent value 
is an important verification of the running of $m_c$, 
one of the basic features of QCD, at the same level as earlier evaluations of 
the running of $m_b$ 
or of the running of $\alpha_s$.

\section*{Acknowledgements}

Part of this work was carried out within the scope of the PROSA, ZEUS and H1 
Collaborations.

\noindent
\begin{flushleft}

\end{flushleft}

\newpage

\begin{table}[!hp]
  \renewcommand{\arraystretch}{1.3}
  \centering
  \begin{tabular}[t]{%
  |c|r|c|r|l|l|}

\hline
       {Subset } & $N_{\rm dat}$ & $Q^{2}$ range &{$\mu\,\,\,\,\,\,$} &
	  {$m_{c}(m_{c})$} & {$m_{c}(\mu)$}  \\
      {} & &{[GeV$^{2}$]} &{[GeV]} &{[GeV]}\ \ fit\ \ \ \ \ scale & 
      {[GeV]} \ \ \ fit  \\ \hline
    1 & 15\ \ &\text{2.5--7\ }& 3.3\ \ & 1.256 $^{+0.078}_{-0.070}$ $^{+0.054}_{-0.000}$ &  0.984 $^{+0.085}_{-0.076}$ \\ 
    2 & 12\ \ & \text{12--18} & 4.5\ \ & 1.192 $^{+0.075}_{-0.073}$ $^{+0.043}_{-0.000}$ &  0.867 $^{+0.077}_{-0.075}$ \\ 
    3 & 13\ \ &\text{32--60}  & 7.0\ \ & 1.208 $^{+0.092}_{-0.088}$ $^{+0.045}_{-0.000}$ &  0.830 $^{+0.089}_{-0.085}$ \\ 
    4 & 7\ \  &\text{120--200}&12.7\ \ & 1.344 $^{+0.130}_{-0.131}$ $^{+0.073}_{-0.074}$ &  0.90 $\pm$ 0.12          \\ 
    5 & 4\ \ &\text{350--650} &21.9\ \ & 1.14\ \ \ $^{+0.22\ \,}_{-0.22}$ $^{+0.13\ \,}_{-0.16}$ &  0.68 $\pm$ 0.19 \\ 
    6 & 1\ \ &\text{2000}     &44.8\ \ & 1.05\ \ \ $^{+0.68\ \,}_{-0.76}$ $^{+0.40\ \,}_{-0.15}$ &  0.56 $\pm$ 0.56 \\ \hline

  \end{tabular}
  \caption{Values of $m_{c}(m_{c})$ at different scales $\mu$, determined 
from six different subsets, and corresponding values of $m_{c}(\mu)$. 
The first uncertainty (fit) corresponds to the uncertainty 
$\delta^{\rm exp}_{\rm fit}$ added in quadrature with the symmetrised systematic 
uncertainties $\delta_1-\delta_6$. 
The second uncertainty (scale) of $m_c(m_c)$ corresponds to the scale variation 
uncertainty $\delta_7$. 
No scale uncertainty is quoted for $m_c(\mu)$ (see text).
The range of $Q^2$ values contributing to the six data 
subsets shown in Fig. \ref{fig:sigred} is given. Also given is the 
corresponding logarithmic average scale $\mu$ for each subset according to 
Eq. (\ref{eq:scale}), and the number $N_{\rm dat}$ of charm data points 
contributing to each measurement.}
 \label{tab:runval} 
\end{table}

\begin{table}[!hp]
  \renewcommand{\arraystretch}{1.3}
  \centering
  \begin{tabular}{|c|c|c|c|c|c|c|c|c|c|c|} 

\hline
 Subset &  $\delta_{\rm fit}^{\rm exp}$ & $\delta_{1}$ & $\delta_{2}$ & $\delta_{3}$ & $\delta_{4}$ & $\delta_{5}$ & $\delta_{6}$ & $\delta_{7}$  \\
        &                             & ($m_b$)      & ($\alpha_s$) & ($f_s$)     & ($Q_0$)  & ($Q^2_{\rm min}$) & (param.)    & (scale) \\
 &  {[\%]} & {[\%]} &{[\%]} & {[\%]} & {[\%]} & {[\%]} & {[\%]} & {[\%]}  \\ \hline

1 	  	&  \, $\pm$ 5.4 & $^{+0.1}_{-0.4}$  & $^{-1.2}_{+2.6}$ & $^{-0.4}_{+0.2}$ & +0.5 & +1.4 & +0.5 & $^{+3.1}_{+4.3}$ \\
2   	&  \, $\pm$ 6.0& $^{+0.2}_{-0.5}$ & $^{-0.9}_{+0.7}$ & $^{-0.5}_{+0.2}$ & +0.3 & +1.0 & +0.9 & $^{+2.4}_{+3.6}$ \\
3 	 	&  \, $\pm$ 7.2 &$^{+0.3}_{-0.7}$ & $^{-0.4}_{+0.3}$ & $^{-0.8}_{+0.3}$ & +1.7 & +0.3 & +1.8 & $^{+0.1}_{+3.7}$ \\
4  	& \, $\pm$ 9.6 & $^{+0.5}_{-0.8}$ & $^{+0.7}_{-0.6}$ & $^{-0.8}_{+0.5}$ & +0.5 & --1.2 & +0.1 & $^{-5.5}_{+5.4}$ \\ 

5  &\, $\pm$ 19.2 	& $^{+0.5}_{-1.2}$ & $^{+1.6}_{-1.8}$ & $^{-1.2}_{+0.5}$ & --0.5 & +2.1 & --1.7 & $^{-14.3}_{+11.6}$ \\
6 	 	& \, $\pm$ 63.8 & $^{-7.4}_{-2.9}$ & $^{+5.9}_{-5.7}$ & $^{-3.0}_{-7.6}$ & +6.5 & --33.3 & +9.5 & $^{+38.1}_{-14.2}$ \\ \hline
  \end{tabular}
  \caption{Summary of the systematic uncertainties in the $m_c(m_c)$ determinations. The 
definitions of the uncertainty sources, the meaning of the symbols in the first and second row and related details are given in the 
text. In cases where opposite variations of 
a variable yield uncertainties with the same sign, only the larger 
one is considered for the uncertainty combination in Table~\ref{tab:runval}. Except for $\delta_7$,
these uncertainties also apply to $m_c(\mu)$, before evolution to the appropriate scale.}
    \label{tab:runsyst}
\end{table}

\newpage
\newpage

\newpage

\newpage

\unitlength1cm

\begin{thebibliography}{99}

\bibitem{PETRA} O. Biebel,
Phys. Rept. {\bf 340} (2001) 165; and references therein.
\\ 
J. Schieck {\it et al.} [JADE Collaboration], 
Eur. Phys. J. {\bf C73} (2013) 2332 [arXiv:1205.3714].\\
S. Bethke {\it et al.} [JADE Collaboration],
Eur. Phys. J. {\bf C64} (2009) 351 [arXiv:0810.1389].

\bibitem{LEPalpharun}
G. Abbiendi {\it et al.} [OPAL Collaboration],
Eur. Phys. J. {\bf C71} (2011) 1733 [arXiv:1101.1470]. \\ 
G. Dissertori {\it et al.},
JHEP {\bf 0908} (2009) 036 [arXiv:0906.3436]. 

\bibitem{HERAalpharun}
S. Chekanov {\it et al.} [ZEUS Collaboration], 
Eur. Phys. J. {\bf C44} (2005) 183 [hep-ex/0502007].
\\  
F.D. Aaron {\it et al.} [H1 Collaboration],
Eur. Phys. J. {\bf C65} (2010) 363 [arXiv:0904.3870]. 
\\
F.D. Aaron {\it et al.} [H1 Collaboration], 
Eur. Phys. J. {\bf C67} (2010) 1 [arXiv:0911.5678].
\\
H. Abramowicz {\it et al.} [ZEUS Collaboration], 
Nucl. Phys. {\bf B864} (2012) 1 [arXiv:1205.6153].
\\
V. Andreev {\it et al.} [H1 Collaboration], 	
Eur. Phys. J. {\bf C77} (2017) 215 [arXiv:1611.03421].

\bibitem{Tevaalpharun}  	 	
V.M. Abazov {\it et al.} [D0 Collaboration], 
Phys. Rev. {\bf D80} (2009) 111107 [arXiv:0911.2710].
\\
V.M. Abazov {\it et al.} [D0 Collaboration], 
Phys. Lett. {\bf B718} (2012) 56 [arXiv:1207.4957].

\bibitem{LHCalpharun}  	
S. Chatrchyan {\it et al.} [CMS Collaboration], 
Eur. Phys. J. {\bf C73} (2013) 2604 [arXiv:1304.7498]. \\
V. Khachatryan {\it et al.} [CMS Collaboration],
Eur. Phys. J. {\bf C75} (2015) 186 [arXiv:1412.1633].\\ 
V. Khachatryan {\it et al.} [CMS Collaboration],
Eur. Phys. J. {\bf C75} (2015) 288 [arXiv:1410.6765]. \\
V. Khachatryan {\it et al.}, [CMS Collaboration],  	
JHEP {\bf 1703} (2017) 156 [arXiv:1609.05331]. 


\bibitem{Baikov:2014qja}
  P.A.~Baikov, K.G.~Chetyrkin and J.H.~K\"uhn,
  JHEP {\bf 1410} (2014) 076
  [arXiv:1402.6611].

\bibitem{beautyrun} J. Abdallah {\it et al.} [DELPHI Collaboration], 
  Eur. Phys. J. {\bf C55} (2008) 525 [arXiv:0804.3883]. \\
  R. Barate {\it et al}. [ALEPH Collaboration], Eur. Phys. J. {\bf C18} (2000) 1 [hep-ex/0008013]. \\
  G. Abbiendi {\it et al.} [OPAL Collaboration], Eur. Phys. J. {\bf C21} (2001) 411 [hep-ex/0105046]. \\
  A. Brandenburg {\it et al.}, Phys. Lett. {\bf B468} (1999) 168 [hep-ph/9905495]. \\
  J. Abdallah {\it et al.} [DELPHI Collaboration], Eur. Phys. J. {\bf C46} (2006) 569 [hep-ex/0603046]. 

\bibitem{ZEUSbeauty}
  H.~Abramowicz {\it et al.} [ZEUS Collaboration],
  JHEP {\bf 1409} (2014) 7 [arXiv:1405.6915].

\bibitem{Addendum} ZEUS collaboration, addendum to \cite{ZEUSbeauty}, unpublished.

\bibitem{Andriithesis}
  A.~Gizhko, {\it Measurement of beauty quark mass at HERA and impact on Higgs production in association with beauty quarks}, 
  DESY-THESIS-2016-015.

\bibitem{HQreview} O. Behnke, A. Geiser, M. Lisovyi, 
  Prog. Part. Nucl. Phys. {\bf 84} (2015) 1 [arXiv:1506.07519]. 

\bibitem{PDG12} J. Beringer {\it et al.} [Particle Data Group],  
Phys. Rev. {\bf D86} (2012) 1.

\bibitem{PDG16} C. Patrignani {\it et al.} [Particle Data Group],  
Chin. Phys. {\bf C40} (2016) 100001.

\bibitem{HERAcharmcomb}
  F.D.~Aaron {\it et al.}  [H1 and ZEUS Collaborations],
  \EPJC {\bf 73} (2013) 2311 [arXiv:1211.1182]; and references therein.

\bibitem{adlm2} 
   S.~Alekhin {\it et al.},
   Phys.\ Lett.\ {\bf B718} (2012) 550
   [arXiv:1209.0436].
   \\
   S.~Alekhin {\it et al.},
   Phys. Lett. {\bf B720} (2013) 172 [arXiv:1212.2355].
   \\
   S. Alekhin, J. Bl\"umlein, S. Moch, 
   Mod. Phys. Lett. {\bf A28} (2013) 1360018 [arXiv:1307.1219].

\bibitem{CTEQmc} J. Gao, M. Guzzi, P.M. Nadolski, 
  Eur. Phys. J. {\bf C73} (2013) 2541 [arXiv:1304.3494].

\bibitem{FONLLmc} V.~Bertone {\it et al.} [xFitter Developers' Team Collaboration],
  JHEP {\bf 1608} (2016) 050 [arXiv:1605.01946].

\bibitem{ABMP16} S. Alekhin {\it et al.}, 
DESY-16-179, DO-TH-16-13, arXiv:1701.05838.

\bibitem{Alekhin:2013nda}
  S.~Alekhin, J.~Bl\"umlein and S.~Moch,
  Phys.\ Rev.\ {\bf D89} (2014)  054028
  [arXiv:1310.3059]; and references therein. \\
  For some of the most recent NNLO developments, see also \\
  J. Ablinger {\it et al.}, Nucl. Phy. {\bf B890} (2014) 48 [arXiv:1409.1135];
  Nucl. Phys. {\bf B882} (2014) 263 [arXiv:1402.0359];
  Phys. Rev. {\bf D92} (2015) 114005 [arXiv:1508.01449];\\
  A. Behring {\it et al.}, Eur. Phys. J. {\bf C74} (2014) 3033 [arXiv:1403.6356].  

\bibitem{DIScomb}
  F.D.~Aaron {\it et al.}  [H1 and ZEUS Collaboration],
  JHEP {\bf 1001} (2010) {109} [arXiv:0911.0884].

\bibitem{DIScombII} 	
  H.~Abramowicz {\it et al.}  [H1 and ZEUS Collaboration],
    Eur. Phys. J. {\bf C75} (2015) 580 [arXiv:1506.06042].


\bibitem{openqcdrad}
  S. Alekhin, {\tt OPENQCDRAD-1.5},\\
  \url{http://www-zeuthen.desy.de/~alekhin/OPENQCDRAD}.
  \\

\bibitem{herafitter}
  HERAFitter, \url{http://herafitter.org}.
  \\
  M.~Botje,
  Comput.\ Phys.\ Commun.\  {\bf 182} (2011) 490
  [arXiv:1005.1481].
  \\
  F.~James and M.~Roos,
  Comput.\ Phys.\ Commun.\  {\bf 10} (1975) 343.

\bibitem{abkm09msbar} 
  S.~Alekhin {\it et al.},
  Phys.\ Rev.\ {\bf D81} (2010) 014032 [arXiv:0908.2766].
\\
  S.~Alekhin and S.~Moch,
  Phys.\ Lett.\ {\bf B699} (2011) 345 [arXiv:1011.5790].

\bibitem{RUNdec} B. Schmidt, M. Steinhauser, 
  Comput.\ Phys.\ Commun.\  {\bf 183} (2012) 1845
  [arXiv:1201.6149].
\\
K.G. Chetyrkin, J.H. K\"uhn, M. Steinhauser,
  Comput. Phys. Commun. {\bf 133} (2000) 43 [hep-ph/0004189].

\end{thebibliography}
\end{document}